\documentclass[12pt]{article}
\usepackage{epsfig,graphicx}

\title{Mass generation with adjoint fermions}
\author{ Yu.A.Simonov\\
State Research
Center\\Institute of Theoretical and Experimental Physics, \\
Moscow, 117218 Russia}
\newcommand{\beq}{\begin{eqnarray}}
 \newcommand{\eeq}{\end{eqnarray}}
\newcommand{\be}{\begin{equation}}
 \newcommand{\ee}{\end{equation}}

 \def\la{\mathrel{\mathpalette\fun <}}
\def\ga{\mathrel{\mathpalette\fun >}}
\def\fun#1#2{\lower3.6pt\vbox{\baselineskip0pt\lineskip.9pt
\ialign{$\mathsurround=0pt#1\hfil ##\hfil$\crcr#2\crcr\sim\crcr}}}

\newcommand{{\SD}}{\rm SD}

\newcommand{\veY}{\mbox{\boldmath${\rm Y}$}}
\newcommand{\vex}{\mbox{\boldmath${\rm x}$}}
\newcommand{\vey}{\mbox{\boldmath${\rm y}$}}

\newcommand{\vesig}{\mbox{\boldmath${\rm \sigma}$}}

\newcommand{\vep}{\mbox{\boldmath${\rm p}$}}

\newcommand{\ves}{\mbox{\boldmath${\rm s}$}}

\newcommand{\veH}{\mbox{\boldmath${\rm H}$}}
\newcommand{\veE}{\mbox{\boldmath${\rm E}$}}

\newcommand{\lan}{\langle}
\newcommand{\ran}{\rangle}

\begin{document}
\maketitle
\begin{abstract}

The QCD-like  gauge theory with adjoint fermions is  considered in the field
correlator  formalism and the total spectrum of mesons  and glueballs is
obtained  in agreement with available lattice data. A new state of a white
fermion appears, as a  bound state of the  adjoint   fermion and gluon  with
the mass close to that of glueball. It is shown, that the main features of
spectra and thermodynamics of adjoint fermions can be explained by this new
bound state.

\end{abstract}

\section{Introduction}

Gauge theories with adjoint quarks are an object of intensive studies both on
the lattice \cite{1}-\cite{6} and in theory \cite{7,8}, see \cite{9} for a
review of theoretical models. This interest is mostly connected with the
technicolor models, which suggest adjoint techniquarks as the source of quark
masses and interaction at high scale \cite{10,11}.

At the same time theories with adjoint quarks present an interesting example of
theories, where one can test mechanisms  developed in the framework of the
Field Correlator Method (FCM)\cite{12} for   confinement \cite{13}, chiral
symmetry breaking (CSB) \cite{14},  and  temperature phase
 transition \cite{15},
known for fundamental fermions .

In particular, lattice data with adjoint fermions \cite{4,5,6} exhibit a
completely different hadron spectra, where  an equivalent of the pion is
heavier, than the glueball.  Moreover, CSB seems to have two different
thresholds in temperature \cite{1}: one, coinciding with deconfinement
temperature  $T_{\rm dec}$, and another much higher, $T_\chi \simeq 6.6 ~T_{\rm
dec}$, where the remnants of CSB disappear.  We shall argue below, that these
features can be explained in the framework of FCM in a simple way.  The basic
element,  as will be shown below is, that  an adjoint fermion can be bound with
a gluon, forming a colorless fermion with nonzero mass, which we call
gluequark. Moreover, also the adjoint fermion  itself can acquire mass,
coupling to gluequark, and this creates a completely different picture of CSB
-- and of fermion mass generation in principle. These both facts can be
important, since this new mechanism is different from  the original
mass-generating mechanism of ordinary quarks, suggested in technicolor
theories.

It is a purpose of the present paper to apply to the  gauge  theory with
adjoint  fermions, (which we call shortly  AdQCD) the Field Correlator Method,
which has provided a selfconsistent mechanism of confinement \cite{13}, and
explained the interconnection of  chiral symmetry   restoration \cite{14}  and
temperature deconfinement \cite{15} in case of fundamental quarks (see
\cite{12} for reviews). Following this method we define an exact path-integral
representation (the so-called Fock-Feynman-Schwinger representation (FFSR))
\cite{16}  for fermions and gluons in the confining background field, which is
characterized by gauge invariant correlators, and derive relativistic
Hamiltonians for white systems of quarks and gluons ($qg)$, two gluons ($gg)$,
and mesons $(q\bar q)$ and baryons ($qqq)$.

This enables us to obtain the corresponding spectra of masses. In doing so we
encounter in the interaction potential $V(R) = V_{\rm conf} (R) + V_1 (R)$ an
interesting phenomenon, which is not known for fundamental fermions in the
quenched approximation: the static potential $V_{\rm conf}$  does not grow
beyond some distance $R_{cr}$ which corresponds to the threshold for creation
of two white fermions -- bound state $qg$. This is similar to the  situation in
the unquenched QCD, where the growth of static potential stops beyond some
distance, but in AdQCD this transition is more sharp, as shown by lattice data
\cite{1,17}. Another unexpected feature -- the strong interaction $V_1(R)$ in
the deconfined phase, which is able to bind quark  and gluon into a white
fermion, as it is done by $V_{\rm conf}(R)$ in the confining phase.

In the last part of the paper we discuss thermal properties of AdQCD and define
$T_{\rm deconf}$ and  show, that it coincides with  the first CSB transition,
which we call $T_{\chi 1}, T_{\chi 1} = T_{\rm deconf}$. However, we show, that
the remnants of mass of the white $(qg)$ state can be nonzero beyond $T_{\chi
1}$ and this part of mass (due to  the nonconfining correlator $D_1$)  produces
chiral condensate, which is gradually decreasing  with temperature, finally
disappearing at $T_{\chi 2}$, which signals a full recovery of chiral symmetry.

Our results may be of importance for technicolor theories (TC), since they
provide a new mechanism of fermion mass generation, and moreover, they allow to
calculate these masses explicitly. The paper is organized as follows: the
section 2 is devoted to the derivation of  the AdQCD Hamiltonian from  FFSR,
for all possible white combinations. In section 3 spectra of all states are
calculated and compared with lattice data.  At the end of the section  the
chiral properties of mesons are discussed and compared to lattice data. In
section 4 the temperature phase transition is discussed for confinement and CSB
phenomena. Section 5  contains summary and perspectives.

\section{Gauge fields with adjoint quarks. Green's functions}

We start with the action for AdQCD in  the Euclidean space-time \be S=\frac14
\int (F^a_{\mu\nu} (x))^2 d^4 x - i \int \psi^+ (\hat D + m) \psi d^4
x,\label{1}\ee where $\psi  $ is  an  adjoint fermion, which can be written
both in double fundamental and adjoint indices \be \psi_{\alpha\beta} = \psi^a
t^a_{\alpha\beta}, ~~\psi^a(x) \to U^+_{ab}(x) \psi^b(x), ~~ \psi_{\alpha\beta}
(x) \to U^+_{\alpha\alpha'}  \psi_{\alpha'\alpha'} U_{\beta'\beta}\label{2}\ee
\be \psi^+_a (D_\mu)_{ab} \psi_b = \psi^+_a ( \partial_\mu \delta _{ab} + g
A_\mu^c f^{abc}) \psi_b .\label{3}\ee

The term $\psi^+\hat D\psi$ can be also be written in fundamental indices,
where gauge invariance is evident, as $$ \psi^+_{\alpha\beta} \hat
D_{\beta\gamma} \psi_{\gamma\alpha} \to U^+\psi^+ U U^+ \hat D U U^+ \psi U =
\psi^+ \hat D\psi.$$

One can now  form white combinations  of $q, g$ as $<$in$|$ and $|$out$>$
hadron states in AdQCD. One can do it  both with the total vector field
$A_\mu$, and (more conveniently) in the background field  formalism \cite{18},
separating vacuum background field $B_\mu,~~ A_\mu =B_\mu + a_\mu$, where gauge
transformations are \be B_\mu \to U^+ (B_\mu + \frac{i}{g} \partial_\mu)U, ~~
a_\mu \to U^+ a_\mu U.\label{4}\ee

In what follows we shall work mostly with the field $a_\mu$, as it was done in
\cite{19,20} for glueballs and gluelumps  respectively, indicating also the
corresponding white combinations with $A_\mu$ (or $F_{\mu\nu} = (\veE, \veH))$.

Fermions $(\psi_{\alpha\beta} $ or $\psi^a$) and gluons
$((A_\mu)_{\alpha\beta}$ or $A_\mu^a$) can be depicted by   thin straight lines
and   by wavy  lines for gluons. In this way one can define mesons, \be \Psi_i
= \psi^+_a \Gamma_i \psi_a = \psi_{\alpha\beta}^+ \Gamma_i
\psi_{\beta\alpha},\label{5}\ee $gg$ glueballs, and $ggg$ glueballs.

\be \Phi_i^{(2)} = tr (a_\mu T_{\mu\nu}^{(i)}  a_\nu), ~~ \Phi_i^{(3)} = tr
(T_{\mu\nu\lambda}^{(i)} a_\mu a_\nu a_\lambda),\label{6}\ee where
$T_{\mu\nu}^{(i)} $, $T^{(i)}_{\mu\nu\lambda}$ contain covariant derivatives,
in simplest cases for $0^{++} gg$ state $T^{(1)}_{ik} = \delta_{ik},
i,k=1,2,3.$

The  equivalent  form for baryons  is $B_i=  f^{abc}\psi^a\psi^b\psi^c$.  There
is however in AdQCD  a gauge invariant object, which is missing in the standard
QCD, namely the $qg$ bound state, which we call gluequark,  \be Q_\mu(x) =tr
(\psi(x) a_\mu(x)),~~ Q^+_\mu(x) = tr (\psi^+(x) a_\mu(x)),\label{7}\ee and
more complicated versions with a matrix operator $\Gamma$ between $a_\mu$ and
$\psi$. The equivalent form can be written with $F_{\mu\nu},$ \be Q^E_i = tr
(\psi(x)E_i(x)), ~~ Q^H_i = tr (\psi (x) H_i(x)).\label{8}\ee

We  are now in  position to write Green's functions for $q, g$ and finally
 for mesons $\Psi_i$, glueballs $\Phi_i$ and gluequark $Q_\mu$.

The first two FFSR for $q,g$ have been written before \cite{16}, and one should
only account for the
 adjoint representation of  the fermion
$$ G_q(x,y) = \lan \psi (x)\psi^+(y)\ran_q = \lan x | (m+\hat D)^{-1}|y\ran=$$
\be =\lan | x (m-\hat D) (m^2-\hat D^2)^{-1}|y \ran = (m-\hat D)\int^\infty_0
ds (Dz)_{xy} e^{-K_q}\phi_q(x,y)\label{9}\ee where $(Dz)_{xy}$ is the path
integral element, $K_q= m^2_qs +\frac14 \int^s_0
\left(\frac{dz_\mu}{d\tau}\right)^2 d\tau,$ and \be \phi_q (x,y)=P_A\exp (ig
\int^x_y A_\mu (z) dz_\mu) P_F \exp (g\int^s_0 d\tau \sigma_{\mu\nu}
F_{\mu\nu}),\label{10}\ee where $P_A, P_F$ are ordering operators and
$\sigma_{\mu\nu} F_{\mu\nu}=\left(\begin{array}{ll}
\vesig\veH&\vesig\veE\\
\vesig\veE&\vesig \veH\end{array}\right)$.

In a similar way one can write the gluon Green's function of the field $a_\mu$
in the background field $B_\mu$ \cite {21} \be G_g (x,y)_{\mu\nu}=\lan x|
(D^2_\lambda\delta_{\mu\nu} - 2ig F_{\mu\nu})^{-1}|y\ran= \int^\infty_0 ds
(Dz)_{xy} e^{-K_g}\phi_{\mu\nu}(x,y),\label{11}\ee where $K_g=\frac14 \int^s_0
\left(\frac{dz_{\mu}}{d\tau}\right)^2 d\tau$ and $\phi_{\mu\nu}$ is \be
\phi_{\mu\nu} (x,y) = {P_A \exp (ig \int^x_y B_\lambda dz_\lambda) P_F \exp (2g
\int^s_0 d\tau F_{\sigma\rho} (z(\tau))}{\mu\nu}\label{12}\ee

Now the Green's function of any white two --component  system $(q\bar q, gg$ or
$qg$)
 averaged over background field $B_\mu$ can be written in the form

\be G_{ik} (x,y) =S \int^\infty_0 ds \int^\infty_0 ds' (Dz)_{xy} (Dz')_{xy}
e^{-K_i-K_k}\lan W_F\ran\label{13}\ee where $ i,k = q\bar q, gg, qg, \bar q g;~
S$ contains possible $(m-\hat D)$ operators and $\lan W_F\ran$ is the Wilson
loop with insertions of operators $F_{\mu\nu}$, \be \lan W_F \ran = tr P_B P_F
\lan\exp\{ ig \int_C B_\mu dz_\mu +\sum_i 2 g \int (s^{(i)} F)
d\tau\}\ran_B.\label{14}\ee

Here $(s^{(q)}F)=\frac12 \sigma_{\mu\nu} F_{\mu\nu}$ for a fermion (of spin
$\frac12$) and $(s^gF)= (\ves^g\veH+ \tilde{\ves}^g\veE)_{\mu\nu} = - i
F_{\mu\nu}$, and gluon spin operators are introduced  as follows \cite{19}

\be (s_m^{(g)})_{ik} = -ie_{mik}, ~~ (\tilde s_m^g)_{i4}
=-i\delta_{im}.\label{15}\ee

The relativistic Hamiltonian can now be derived from (\ref{13})  as in
\cite{22}, assuming smooth trajectories for  both partners in $G_{ik}$ (which
actually implies  that  corrections from  $Z$ graphs with backward-in-time
motion are included in the selfenergy terms, explicitly written). In this way
one obtains (see \cite{22}, \cite{23} for details of derivation) \be ds
(D^4z)_{xy} \to (D^3 z)_{\vex\vey} \frac{D\omega}{2\bar\omega},\label{16}\ee
and $\bar\omega$ is the average quark (or gluon) energy inside hadron (the bar
sign over $\omega$ will be omitted in what follows). In this way, following
\cite{22}, \cite{23} one obtains both hadron coupling constant $f_h$, and the
relativistic Hamiltonian in the c.m. system, which we shall write for
simplicity for the zero angular momentum as a common form in three cases: a)
for mesons; b) for gluballs;  c) for gluequark

\be
 H = \sum_i \sqrt{\vep^2 + m^2_i}+ \sigma_{\rm adj} r + V_c^{\rm (adj)} (r) + V_{ss}
 \label{17a}\ee and its einbein form \cite{22}

\be H_\omega=\sum_i\left(\frac{\omega_i}{2} +
\frac{\vep^2+m^2_i}{2\omega_i}\right)+ \sigma_{\rm adj} r +V^{\rm (adj)}_c (r)
+ V_{ss}\label{17}\ee where $V_c^{\rm (adj)}(r)$ is an effective gluon exchange
potential, for gluon $m_i=0$ and $V_{ss}$ is the spin-dependent interaction,
and
 we shall be interested below in $L=0$ states and keep only the
hyperfine interaction $V_{ss}$, which is slightly modified for adjoint sources
\cite{19} (gluon exchanges in $V_c^{\rm (adj)}$ are strongly reduced by BFKL
loop corrections, see \cite{19} for details) \be V_{ss}
=\frac{\ves^{(i)}\ves^{(j)}}{3\omega_i\omega_j} V_4(r),~~ V_4(r) = 5\pi C_2
({\rm adj}) \alpha_s \delta^{(3)} (r) \label{18}\ee

Here $\omega_i$ in the einbein approximation \cite{22} (better than 5\% for
lowest states) is found from the condition on the resulting mass $M$, $
H_\omega \Psi = M\Psi,$ \be \frac{\partial M (\omega_1, \omega_2)}{\partial
\omega_i} =0, ~~ i=1,2.\label{19}\ee

The  mass  $M$ -- the eigenvalue of the Hamiltonian (\ref{17}) -- can be
written as

\be M_n= \sum^i_{i=1} \frac12 \left(\frac{m^2_i}{\omega_i}+\omega_i\right) +
\varepsilon_n (\tilde \omega)+ \Delta M_{ss}, ~~ \tilde \omega=
\frac{\omega_1\omega_2}{\omega_1+\omega_2},~~ \label{20}\ee$$\Delta M_{ss} =
\frac{5\alpha_s \sigma_{\rm adj}}{2(\omega_1 + \omega_2)} (\ves_i \ves_j)$$
where $\varepsilon_n$ is expressed via dimensionless numbers $a_n$ \be
\varepsilon_n = (2\tilde \omega)^{-1/3}(\sigma_{\rm adj})^{2/3}
a_n.\label{21}\ee

We shall be interested in the lowest states $a_0 =2.338,~~ a_1=4.088$ for $L=0$
and $n_r=0$ and 1 respectively.

Eq. (\ref{19}) with the help of Eqs. (\ref{20}), (\ref{21}) yields $\omega_1,
\omega_2$ \be \omega^2_i = m^2_i + \frac{(\sigma_{\rm adj})^{2/3}a_n(2\tilde
\omega)^{2/3}}{3}, ~~ i=1,2.\label{22}\ee

For $m_1=m_2=0$ one obtains $\omega= \left(\frac{a_n}{3}\right)^{3/4}
~~\sqrt{\sigma_{\rm adj}},$ while for heavy quarks, $m_i\>> \sqrt{\sigma_{\rm
adj}},$ one has\be \omega^2_i \approx m^2_i+\frac{(\sigma_{\rm adj})^{2/3}a_n
\tilde m^{2/3}}{3} +... .\label{23}\ee

Finally, the hyperfine correction $\Delta M_{ss}$ is \be \Delta M_{ss} =\frac{
5N_c}{6}  \frac{\alpha_s (hf) \sigma_{\rm adj}}{(\omega_1+\omega_2)}
\ves_1\ves_2, \label{24}\ee
$$\ves_1\ves_2=\frac{J(J+1)-s_1(s_1+1)-s_2(s_2+1)}{2},$$ where $J, s_1, s_2$ are
spins of the bound system and of its components respectively, $\alpha_s(hf)$ is
the effective $\alpha_s$ in the hyperfine interaction, we take it $\alpha_s
(hf)=0.25$, and we neglect the nonperturbative part of hyperfine interaction,
see \cite{19} for a discussion.

One should add a short discussion on $V_c^{\rm (adj)} (r)$.

Naively one could assume, that gluon exchange potential $V_c(r)$ between
adjoint quarks or gluons is the same, as between fundamental quark and
antiquark, but multiplied by the  Casimir factor $9/4$. However, one can argue,
as it was done in \cite{19}, that such strong interaction is largely reduced
due to formation of the quark-gluon chains of BFKL type, which produces a
rather weak resulting interaction, which can be deduced from the relatively
small shift $\Delta = \alpha_P (0) -1$ of the intercept. This conclusion is
also confirmed by lattice calculations in \cite{24}, where the resulting masses
for glueballs do not show significant contribution from $V_c(r)$. Therefore we
neglect $V_c(r), \Delta M_c$ in the first approximation and keep perturbative
gluon contribution only in the spin-dependent interaction $V_{ss}$.

Looking at the Hamiltonian (\ref{17}) in the case of zero fermion current mass
$m_i$, one can see, that the resulting mass eigenvalues in the systems $gg$ and
$qg$ may differ for the zero quark mass only due to hyperfine interaction,
which is large  in both systems due to adjoint Casimir  factor $9/4$. E.g. in
lattice calculations the mass difference between $2^{++}$ and $0^{++}$
glueballs is around 0.7 GeV.

We are now in position to calculate  the glueball and gluequark  bound states,
using the Hamiltonian (\ref{17}), (\ref{18}) with the condition (\ref{19}). The
result for glueballs are essentially the same as in \cite{19}, here we only
slightly change $\alpha_s(hf)$ in the $hf$ interaction (\ref{18}), taking in
this paper $\alpha_s^{(eff)}=0.25$. Then the difference  between $qg$ and $gg$
bound states is only in the $ss$  term due to different spin  values: one
obtains total spin value $J=0,2$ for $gg$ and $J=\frac12,~\frac32$ for $qg$
states.

Results are given in the Table 1 for $\sigma_f=0.18$ GeV$^2$ and assuming zero
masses of adjoint quarks.

\begin{table}[h]
\caption{ The $L=0$ masses of $gg$ and $qg$ bound states,
$\frac{m_{gg}}{\sqrt{\sigma_f}}, $ $\frac{m_{qg}}{\sqrt{\sigma_f}} $ for
$N_c=3,~ \alpha_s(hf) =0.25$ in comparison with lattice data
\cite{24}.}\begin{center}
\begin{tabular}{|c|r|r|r|r|}\hline

System& \multicolumn{2}{c|}{$gg$}&\multicolumn{2}{c|}{$qg$}\\\hline

$J^P$& $0^+$& $2^+$& $\frac12^{-}$& $\frac32^{-}$\\\hline $m/\sqrt{\sigma_f}$,
~~this work& 3.56&5.30&4.15&5.02\\\hline $m/\sqrt{\sigma_f}$, $SU(3)$ [24] &
3.55&4.78&-&-\\\hline

 $m/\sqrt{\sigma_f}$, $SU(2)$ [24] & 3.78&5.45&-&-\\\hline

\hline
\end{tabular}
\end{center}
\end{table}
%%%%%%%%%%%%%%%%%%%%%%%%%
%%%%%%%%%%%%%%%%%%%%%%%%%%%%%
%%%%%%%%%%%%%%%%%%%%%%%%%
 The r.m.s. radii of these bound states are around $R_0\approx (0.3\div 0.4) $ fm
  and hence one can expect, that for $R\ga  R_{cr}  \approx (0.6\div 0.8)$ fm
 the total potential between adjoint quarks does not grow and becomes flat. Qualitatively this  picture agrees with what
was found on the lattice in the $SU(3)$ theory with adjoint fermions \cite{1},
\cite{17}.

Of special interest are vector $V$ and pseudoscalar PS meson masses, which can
be computed from the  current quark masses (assumed here to be vanishingly
small), or for the quark masses equal to $m(qg)$. The resulting masses for
$m_q=0$  are expressed in  terms of $\sigma$ and were computed  in
\cite{22}-\cite{25}. (Note, that the selfenergy correction \cite{26} is
important here). Results for $L=0,~~n=0$ with the assumption of zero quark mass
are easily found from (\ref{20}),(\ref{24}) (here we do not exploit the
additional Nambu-Goldstone mechanism of PS mass suppression, as was done in
\cite{14,25}, which would drive $m_{\rm PS}$ to zero for zero quark mass). One
obtains  for adjoint quark-antiquark states \be m_{\rm PS} (m_q=0) =0.685 ~{\rm
GeV},~~ m_V = 1.075~{\rm GeV}.\label{25}\ee

These values are in strong disagreement with lattice data for $n_f =2$ in
\cite{2,4}, where \be \frac{M_{PS}}{\sqrt{\sigma_{\rm adj}}} \approx 7, ~~
\frac{M_V}{M_{PS}} \approx 1.04.\label {26}\ee

Therefore we shall explore now another dynamics, where the bare quark mass
coincides with $m_{qg} \left(\frac12^-\right)$. As one can  see in Table 1,
$m_{qg}\left(\frac12^-\right)\cong 2.8 \sqrt{\sigma_{\rm adj}}$, and the
average energy of the effective $(q\bar q)$  can be obtained from (\ref{22},
(\ref{23}), which reduces to the equation for $x=\left( \frac{\sqrt{\sigma_{\rm
adj}}}{\omega}\right)^{2/3}$: $ 1=\frac{m^2_{qg}}{\sigma} x^3 + \frac{a_0}{3}
x^2$. Here $\Delta M_{ss} = \Delta_{ss} \sqrt{\sigma_{\rm adj}}~ \ves_1\ves_2$,

\be M_{PS} (M_q= m_{qg}) \cong 2 m_{qg}+ \frac{a_0 \sigma_{\rm
adj}^{2/3}}{m_{qg}^{1/3}} - \Delta_{ss} \frac34 =(7.25- \frac34 \Delta_{ss})
\sqrt{\sigma_{\rm adj}},\label{27}\ee

\be M_V (M_q= m_{qg})=2 m_{qg}+ \frac{a_0 \sigma_{\rm adj}^{2/3}}{m_{qg}^{1/3}}
+ \Delta_{ss} \frac14 =(7.25+ \frac14 \Delta_{ss}) \sqrt{\sigma_{\rm
adj}}.\label{28}\ee

One can assume, that the hyperfine interaction occurs at very small distances,
when both quark and antiquark are bare, i.e.  without clouds of gluons, which
turn them into $(qg)$ and $(\bar qg)$ respectively, i.e. they have
$\omega_q^{(hf)} \approx \omega_g^{(hf)}=\left (\frac{a_0}{3}\right)^{3/4}
\sqrt{\sigma_{\rm adj}} =0.83 \sqrt{\sigma_{\rm adj}},$ and    $ \Delta_{ss} =
\frac{N_c}{3} 0.393  \left( \frac{\alpha_s}{0.25}\right)$.

Then  $\Delta_{ss} = \Delta_{gg} =0.262$   for $\alpha_s (hf) =0.25$,  $N_c=2$
and the   masses are $ M_{PS} \cong 7.05 \sqrt{\sigma_{\rm adj}}, ~~ M_V\cong
7.31 \sqrt{\sigma_{\rm adj}}$, while the ratio is

\be\frac{M_V}{M_{PS}}  = 1.037, ~~ \label{29}\ee

\begin{table}[h]
\caption{ Masses of  $PS$ and $V$  states of $q\bar q$,
$\frac{M_{PS}}{\sqrt{\sigma_{\rm adj}}}, ~~\frac{M_V}{\sqrt{\sigma_{\rm adj}}}$
and their ratio $\frac{M_V}{M_{PS}}$ for $\alpha_s =0.25$ and
$N_c=2$.}\begin{center}
\begin{tabular}{|c|r|r| }\hline

& this work &[4,5]\\\hline
 $\frac{M_{PS}}{\sqrt{\sigma_{\rm adj}}}$& 7.05&$\approx
7$\\\hline $ \frac{M_V}{\sqrt{\sigma_{\rm adj}}}$ &7.31&$\approx 7$\\\hline
$\frac{M_V}{M_{PS}}$& 1.037& 1.04\\

\hline

\hline
\end{tabular}
\end{center}
\end{table}
%%%%%%%%%%%%%%%%%%%%%%%%%

The  resulting values of $M_{PS}, M_V$ are  given in Table 2 in comparison with
approximate values , deduced from  \cite{4,5}. In a similar way one can
consider mesonic states made from the gluequarks of spin $\frac32$. In this
case $M_{ss}=0.075 \sqrt{\sigma_{\rm adj}}\ves_1\ves_2$ and  one has for the
lowest and highest spin states

\be \frac{M_{PS}(J=0)}{\sqrt{\sigma_{\rm adj}}}= 8.30 - \frac{15}{4} \cdot
0.075 = 8.02\label{30}\ee

\be \frac{M(J=3)}{\sqrt{\sigma_{\rm adj}}} =8.30 +\frac94\cdot 0.075 =
8.47\label{31}\ee with the ratio $\frac{M(J=3)}{M_{PS}} = 1.056$.

At this point one must consider the  general picture of CSB and distinguish two
possible
 mechanisms of CSB:

\begin{enumerate}
\item  CSB due to confinement without fermion mass generation, valid for zero
or small fermion  mass generation, $m<m_{crit}$. \item CSB due to large fermion
mass, $m>m_{crit}$.

\end{enumerate}

The difference between these two pictures lies in the mechanism of
Nambu-Goldstone meson creation,  which is valid in the case 1, and is absent in
the case 2.

In our case the appearing fermion mass is the mass of $qg$ state, which is
large    and therefore  one can assume that no Nambu-Goldstone phenomenon  can
take place at our scale.

\section{Thermodynamics of adjoint fermions}

Thermodynamics of QCD with adjoint fermions was studied in \cite{1},\cite{2},
where two  phase transitions
 were found with  $T_{\rm chiral}$ and $  T_{\rm deconf}$, and  $T_{\rm chiral}/T_{\rm deconf} \cong 6.65$ \cite{1}.
 We shall discuss possible features of
thermodynamics of AdQCD, using the formalism of FCM and we show that there are
two chiral transitions, one at $T_{\chi 1} = T_{\rm deconf}$, and another,  at
$T_{\chi 2} = T_{\rm chiral}$.

In this framework the colorelectric field correlators are defined in a gauge
invariant way as

\be \frac{g^2}{N_c} tr (E_i (x) \Phi(x,y) E_k (y) \Phi(y,x)) = \delta_{ik} (D^E
(u) + D_1^E (u) + u^2_4\frac{\partial D_1^E} {\partial u^2_4})+ u_iu_k
\frac{\partial D_1^E}{\partial u^2}, ~~ u \equiv x-y\label{4.1}\ee give rise
inside Wilson loop $\lan W_F\ran$ (\ref{14}) to two interactions between $q$
and $\bar q$, $ q$ and $g$ , or $g$ and $g$ at the temperature $T$

\be V^E (r,T) = 2 \int^r_0 (r-\lambda) \int^{1/T}_0  d\nu (1-\nu T) D^E
(\sqrt{\lambda^2+\nu^2}) \approx \sigma r (r\to \infty)\label{4.2}\ee

\be V^E_1 (r,T) =   \int^{1/T}_0  d\nu (1-\nu T) \int^r_0\lambda d \lambda
D^E_1 (\sqrt{\lambda^2+\nu^2}) = const   (r\to \infty)\label{4.3}\ee

In our previous discussion these potentials were referred to as $V_{\rm conf}$
and $V_1$ respectively.

The physical picture of deconfinement, was suggested in \cite{27} and further
developed in \cite{28},\cite{15}, (see\cite{29} for
 a review),  where agreement with   lattice calculations  is demonstrated.

 It is based on the notion, that $D^E$ (and hence $\sigma$) vanishes at $T\geq
T_{\rm deconf}$. Moreover, this vanishing happens due to the fact, that  the
minimum of the total free energy $F=-P$(or maximum of the total pressure $P$),
consisting of vacuum energy density of gluonic fields
 and the  free energy (pressure) of valent mesons, glueballs, or quarks and gluons,   requires the confining    part of vacuum energy density to
 vanish above $T=T_{ \rm deconf}$, as was found in lattice calculations \cite{30}.

 One can write  equality of $P$ in two phases, $P_I =P_{II}$, where
\be P_I = |\varepsilon_{\rm vac}^{\rm conf} | + \chi_1 (T), ~~ P_{II} =
|\varepsilon_{\rm vac}^{\rm deconf} | +P_{ql} + P_q\label{4.4}\ee Here $ \chi_1
(T),$ is the hadronic  gas pressure, which will be neglected with 10\%
accuracy. We also take quark mass equal to zero.

The gluonic energy density of the vacuum, expressed via gluonic condensate
$G_2$ is

\be \varepsilon_{ \rm vac} =\frac{\beta_0}{32} G_2,  ~~  G_2 = \frac{2\alpha_s
\lan E^2_i + H^2_i\ran_{\rm vac}}{ \pi} = \frac{3N_c}{\pi^2} (D^E(0)+D_1^E(0)
 +D^H(0)+D_1^H(0))\label{4.5}\ee

Taking into account, that $D^E_1(x)$ vanishes for $x\to 0$, and does not
contribute to $G_2$ \cite{31} and the  fact, that  at $T=0$, $D^E(0)=D^H(0)$
(and we keep this equality for $T=T_{\rm deconf}$) one obtains \be
|\varepsilon^{ \rm conf}_{\rm vac}| \approx 2 |\varepsilon_{\rm vac}^{ \rm
deconf}|\label{4.6}\ee

As a result, Eq. (\ref{4.4}0 yields for the transition temperature  $T_c$

\be T_c \equiv T_{ \rm deconf} = \left(\frac{|\beta_0| G_2}{64
(p_g+p_q)}\right)^{1/4}\label{4.7}\ee
$$ p_i = P_{i}/{T^4}, ~~ i=g,f,a$$

Eq.(\ref{4.6}) is applicable to $SU(N_c)$ theories with $n_f$ fundamental and
$n_a$ adjoint quarks, and

\be \beta_0 = \frac{11}{3} N_c - \frac23 n_f - \frac43 N_c n_a.\label{4.8}\ee
In the leading approximation  of the Vacuum Dominance Approach \cite{15} the
pressure $p_i$ is expressed via Polyakov lines and  was calculated in \cite{15}
\be p_g = \frac{2(N_c^2-1)}{\pi^2} L_a, ~~ p_f = \frac{4N_cn_f}{\pi^2}  L_f, ~~
p_a = \frac{4(N^2_c-1) n_a}{\pi^2} L_a\label{4.9}\ee where Polyakov lines $L_a,
L_f$ are expressed via $V_1 (r,T)$\be L_f = \exp \left( - \frac{V_1(\infty,
T)}{2T}\right), ~~
 L_a = \exp \left( - \frac{9V_1(\infty, \pi)}{8T}\right),\label{4.9}\ee

 To  find $V_1(\infty, T)$ one can into account, that  $D_1(x)$ and hence
 $V_1(r,T)$ are expressed via the gluelump with mass $O(1$ GeV) and  were calculated in \cite{31, 32} in
 comparison to lattice  data \cite{33}, yielding approximately  for fundamental quarks \be   V_1
 (\infty, T)=\frac{0.17{\rm~GeV}}{1.35\left(\frac{T}{T_c}\right)-1};~~
 V_1(\infty,
 T_c)= 0.5 {\rm ~ GeV}.\label{4.10}\ee

 Taking  the same value of $V_1(\infty , T_c)\approx 0.5$  GeV in the
  case  of adjoint quarks, one obtains from (\ref{4.7}) for $N_c=3, ~~ n_a =2$
 and $n_f =0$ (as in \cite{1}), and for zero quark masses and $G_2 =0.005$ GeV$^4$ \cite{34},
  assuming free quarks  (with $V_1$
 taken into account in $L_a$)
 \be \exp \left( -\frac{9V_1(\infty)}{32
 T_c}\right) T_{c} = 0.0733~{\rm GeV}, ~~ T_{c} = 0.167~{\rm
 GeV}.\label{4.12}\ee

 On the other hand, if all quarks  above $T_c$ are bound with gluons, the
 resulting mass of gluequark can be around $\frac94 V_1 (\infty, T_c)$ and the
 corresponding contribution to the free energy (pressure) is strongly reduced.
 Neglecting this contribution, one obtains $T_c$ the same, as  in the quenched QCD.

 To check accuracy of this  prediction, we shall calculate $T_c$ for  the pure gluon case and the case with $n_f=3;~ n_a
 =0$. The results are given in Table 3 in comparison to lattice calculation from
 \cite{35}-\cite{38}.

 One can see in Table 3 a
   reasonable agreement with lattice data  for the deconfinement transition in
   case of $m_q=0$.

 Now the connection between confinement and CSB in $SU(N_c)$ theories with
 fundamental quarks was established in \cite{14}, where it was shown, that
 confinement is a scalar interaction and produces effective scalar quark mass
 operator.

 Moreover, in \cite{15} both $f_\pi$ and $\lan \bar q q\ran$ have been
 calculated in terms of $\sigma$ in  good agreement with lattice and
 experiment.

 It was argued in \cite{14}, that the confinement creates in the Green's
 function of the quark  a scalar mass term $S_q(x,y) = (\hat \partial+ \hat{
 \mathcal{M}}(x,y))^{-1}, \hat{\mathcal{M}}(x,y) = \sigma |\vex -\veY|$, where
 $\bar Y$ is an averaged antiquark position (exact in the case of heavy-light
 mesons). Therefore, CSB occurs automatically in the confinement phase and
 disappears exactly  for fundamental quarks at $T_{\rm deconf}$, when  the string tension $\sigma$ is
  identically zero, provided that quark mass is small. Note, that for fundamental
  quarks the bound states  with gluons are impossible, however for adjoint quarks there
   appears another possible source of CSB, namely the possibility of bound $qg$ system due to nonvanishing
   of $D_1(x)$ and hence nonzero  $V_1(r,T)$, which we can extract  from the Polyakov
   loops $L_a, L_f$. To analyze the $qg$ system in this case one must consider
   the relativistic hamiltonian (\ref{17}), but now with replacement of
   $\sigma_{\rm adj} r$ by $V_1(r,T)$

   \be H_{dec} = 2\sqrt{\vep^2}+ \frac94 V_1(r, T)\label{4.13}\ee

   One can estimate  approximately $V_1(\infty, T)$ for different $T$, using
   Polyakov loops measurements   of $L_8, L_3\equiv L_a, L_f$ in \cite{1},
   which are done at several values of $\beta_0 =1/6 g^2  $. Finding  the
   correspondent $T$ values from the two-loop $\beta$ function, one obtains
   that $V_1 (\infty, T) $ from $L_8,L_3$ in \cite{1} is growing with $T$
   roughly proportionally to $\sqrt{\sigma_s}: ~ V_1 (\infty, T) \cong 2
   \sqrt{\sigma_s (t)}$, where $\sigma_s (t)$ is the spacial string tension for
   fundamental quarks. Therefore $\frac94 V_1 (\infty, T) \approx 1.3$ GeV for
   $T\approx 1.8 ~T_{\rm deconf}$  and $\frac94 V_1 (\infty, T)\approx 3$ GeV
   for $T \cong 5.3~ T_{\rm deconf}$. At $T=T_c$ one  has $L_3 =0.2$, which
   yields $V_1(\infty, T_c)\cong 0.55$ GeV in good agreement with \cite{28}.

   Having in mind, that the range of potential $V_1 (r, T)$ is defined by the
   gluelump mass in $D_1 (x)$ which is of the order of 1  GeV, one can expect,
   that the interaction in the Hamiltonian (\ref{4.13}) is strong enough to
   produce a bound state of gluequark also for $T> T_{\rm deconf}$. Variational
   solutions of the equation $H_{\rm dec} \Psi _{qg} = m_{qg} \Psi_{qg}$ also
   support this expectation. A similar situation was observed in the case of
   white bound states of fundamental quarks or gluons above $T_{\rm deconf}$,
   see \cite{32}, \cite{33}, \cite{39}. Leaving details of calculations of AdQCD thermodynamics to
   the future, we finish here with few remarks on the behavior of
   the chiral condensate $\lan \bar \psi_a\psi_a\ran$, which was measured in
   \cite{1},\cite{2} in the wide range of temperatures.

   At this point one should distinguish two sources of CSB and the chiral
   condensate $\lan \bar \psi \psi\ran, $ as was discussed at the end of the
   previous section. In general, the chiral condensate is nonvanishing both in
   the limit of small quark mass $(m_q < m_{\rm crit})$, when  spontaneous CSB due to
   confinement  takes place, and  also due to nonzero  quark mass, which  can
   be dynamically created, e.g. due to $V_1(r,T)$ as discussed above.  It  is
   conceivable, that for  large quark mass (whatever is the mass
   generation mechanism) the quark condensate is also nonzero.

   Qualitatively, one can estimate the behavior of the chiral condensate as a
   function  of the large mass of  a quark $m_q$, as it was done in \cite{34},
   $\lan \bar \psi \psi \ran \approx - \frac{G_2}{12 m_q}$. One can expect,
   that the effective quark mass $m_q$ is connected to $m_{qg}$ and is growing
   with $T$ as $V_1 (\infty, T)$, and hence $\lan \psi \psi\ran$ tends to zero
   with growing $T$ due to large $m_{qg}$, and due to the fact that bound $(qg)$ dissociate with growing temperature.

\begin{table}[h]
\caption{ Transition temperatures in QCD with fundamental  or adjoint quarks
from Eq. (\ref{4.7}) in comparison with lattice data from \cite{35}-\cite{38}.}

 \begin{center}
\begin{tabular}{|l|l|l|l|l|l|l| }\hline

$N_c$& $n_f$& $n_a$& $T_c$(MeV) & $T_c$(MeV) & $T_c$(MeV) & $T_c$(MeV) \\
&&& this work &&[37]&[38]\\\hline 3&0&0&260&269 [35]&&\\\hline
3&2&0&180&$173\pm
8$ [36]&&\\\hline 3&3&0&170&$154\pm9$ [36] & 164(6) &165(5)(3)\\
&&&&&&147(2)(3)\\\hline 3&0&2&167&&&\\\hline

\hline
\end{tabular}
\end{center}
\end{table}

Note, that the potential $V_1(r,T)$ is vector like \cite{40}, hence it cannot,
in contrast to $V_{\rm conf}$, produce CSB by itself, and  only after it is
embedded in the resulting ($qg$) mass, which is scalar, this mass breaks chiral
symmetry.

Finally few remarks about the so-called PCAC mass $m$, defined as in \cite{3,4}
via the ratio of correlators of axial $A$ and pseudoscalar $P$ currents. \be
m(t) = \frac14 [(\partial_0 + \partial_0^*) f_{AP} (t) ] / f_{PP}
(t).\label{47}\ee

As was discussed in \cite{14}, \cite{25}, $m(t)$ can have two regimes, which
are seen in the  asymptotics in two limiting cases: i) $t\la  \lambda$; ii)
$t\gg \lambda$, where $\lambda$ is the vacuum correlation length. The latter is
defined in FCM \cite{12, 30, 31}, $ \lambda\approx 0.1 \div 0.2$ fm. In the
regime  i) $m(t) \approx \lambda \sigma$, while in the case ii) $m(t\gg
\lambda) \approx \sigma \lan r\ran\approx 2.5 \sqrt{\sigma}$, where $\lan
r\ran$ is the average size of the meson estimated for linear confinement. This
latter large scale regime with $\sqrt{\sigma} a \approx 0.4 ma$ was seemingly
observed in the lattice calculations (cf. Fig. 3 of \cite{5}). In the low,
scale case $m(0) = \sigma \lambda (\cong 0.15$ GeV for fundamental quarks)
defines small resulting values of $f_\pi, f_K$ as shown in \cite{25}. Note,
that all our discussion above in the present paper refers to large scale
dynamics only.

\section{Conclusions}

We have studied ADQCD in both confined and deconfined regions, using FCM and
applying relativistic hamiltonian \cite{22} to calculate spectrum of bound
states. In doing so we have discovered   new bound states of quark and gluon
$(qq)$ or gluequarks with masses close  to those of glueballs. We have shown,
that the appearance  of these white fermions drastically changes the whole
spectrum and produces in particular large masses of $PS$ and $V$ mesons, with
the ratio close to unity. These features, and the absolute values  of meson
masses are in close agreement with lattice data. We have also calculated  the
deconfinement temperature in AdQCD, which appeared of the same order as in QCD
with fundamental quarks, and we have also checked $T_{\rm deconf}$ in the
latter $vs$ lattice data. We have argued, that the $qg$ bound state may survive
above $T_{\rm deconf}$ and simulate the nonzero quark condensate, while
dissociation of $qg$ at larger $T$ may provide restoration of chiral symmetry.
The search for gluequarks on the lattice seems highly desirable.

The appearance of massive gluequarks presents itself as a new source of a gauge
invariant  mass  generation mechanism, which can be used at high scale, e.g. in
the framework of TC  and  ETC theories \cite{10, 11}.

The author is grateful for useful discussions to A.M.Badalian and
M.I.Polokarpov.


\begin{thebibliography}{99}

\bibitem{1}F.Karsch and M.Luetgemeier Nucl. Phys. Proc. Suppl. {\bf 73}, 444 (1999);
hep-lat/9809056; Nucl. Phys. B {\bf 550}, 449 (1999); hep-lat/9812023.


\bibitem{2}  J.Kogut, J.Polonyi, D.K.Sinclair and H.W.Wyld, Phys. Rev.  Lett. {\bf 54},1980 (1985);\\ J.Engels, S.Holtmann and T.Schulze, Nucl. Phys. B {\bf 724},
357 (2005);\\ G.Cossu, M.D'Elia, A.Di Giacomo, G.Lacagnina and C.Pica, Phys.
Rev. D {\bf 77},  074506 (2008).
\bibitem{3}
L.Del Debbio, B.Lucini, A.Patella, C.Pica and A.Rado, Phys. Rev. D {\bf 80},
074507 (2009); arXiv: 0907.3896 [hep-lat];\\
 L.Del Debbio, B.Lucini, A.Patella, C.Pica and A.Rado, Phys. Rev. D
{\bf 82}, 014510 (2010); arXiv: 1004.3206 [hep-lat].



\bibitem{4} L.Del Debbio, B.Lucini, A.Patella, C.Pica and A.Rado, Phys. Rev. D
{\bf 82}, 014509 (2010); arXiv: 1004.3197 [hep-lat].





\bibitem{5}  A.Patella, L.Del Debbio, B.Lucini, C.Pica and A.Rado, PoS Lattice 2010:068, 2010; arXiv: 1011.0864[hep-lat].



\bibitem{6} A.J.Hietanen, J.Rantaharju, K.Rummukainen and K.Tuominen, JHEP {\bf 05}, 025 (2009); arXiv:0812.1467
[hep-lat].





\bibitem{7} F.Sannino and  K.Tuominen, Phys. Rev.  D {\bf 71}, 051901  (2005).

%%%%%%%%%%%%%%%%%%%
\bibitem{8} D.D.Dietrich and F.Sannino, Phys. Rev.  D {\bf 75}, 085018 (2007), hep-ph/0611341; S.Catterall and F.Sannini, Phys. Rev.
D {\bf 76}, 034504 (2007) arXiv: 0705.1664  [hep-lat].



\bibitem{9} F.Sannino, arXiv: 0911.0931 [hep-ph].

\bibitem{10} C.T.Hill and E.H.Simmons, Phys. Rept. {\bf 381}, 235 (2003);
hep-ph/0203079.


\bibitem{11} K.Lane, arXiv: hep-ph/0202255.








\bibitem{12}A.Di Giacomo, H.G.Dosch,  V.I.Shevchenko, and Yu.A.Simonov, Phys.
Rept., {\bf 372}, 319 (2002), hep-ph/0007223; Yu.A.Simonov,  in QCD:
Perturbative of  Nonpertubative, L.S. Fereira, P.Nogueira and J.I.Silva-Marcos
eds.; World Scientific,  Singapore, 2001; hep-ph/99112337.

\bibitem{13}H.G.Dosch,  Phys. Lett. B {\bf 190}, 177 (1987);\\ H.G.Dosch   and Yu.A.Simonov, Phys.
Lett., B {\bf  205}, 339 (1988);\\ Yu.A.Simonov, Nucl. Phys. B {\bf 307}, 512
(1988);  D.S.Kuzmenko, V.I.Shevchenko, and Yu.A.Simonov, Phys. Uspekhi {\bf
47}, 1 (2004); hep-ph/0310190.

\bibitem{14}  Yu.A.Simonov, Phys. At. Nucl. {\bf 60}, 2069 (1997); ibid. {\bf
67} 846 (2004); Phys, Rev. D {\bf 65}, 094018 (2002); hep-ph/0201170.



\bibitem{15}Yu.A.Simonov,  Phys. At. Nucl. {\bf 58},  309 (1995), hep-ph/9311216;\\
Yu.A.Simonov,  Ann. Phys.  (N.Y.) {\bf 323 },  783 (2008); hep-ph/0702266;\\
 E.V.Komarov and Yu.A.Simonov,  Ann. Phys. (N.Y.) {\bf 323},1230 (2008); arXiv:0707.0781 [hep-ph].






\bibitem{16}Yu.A.Simonov, Nucl. Phys.  B {\bf 307 }, 512 (1988), Yu.A.Simonov, and J.A.Tjon,  Ann. Phys. (N.Y.) {\bf
300}, 54 (2002);hep-ph/0205165.

\bibitem{17} P.de Forcrand and O.Philipsen, Phys. Lett. B {\bf 475}, 280
(2009);  hep-lat/9912950.
\bibitem{18}
 B.S.De Witt, Phys. Rev. {\bf 162}, 1195 91967); ibid. 1239 (1967);\\
J.Honerkamp, Nucl. Phys. B {\bf 48}, 269 (1972);\\G.'tHooft, Nucl.
Phys. B {\bf 62}, 44 (1973);\\ L.F.Abbot, Nucl. Phys. {\bf 185}, 189 (1981);\\
Yu.A.Simonov,Phys. At.Nucl. {\bf 58}, 107  (1995); Yu.A.Simonov, in: Lecture
Notes in Physics, v. 479, p. 144, Springer, 1996.
\bibitem{19}A.B.Kaidalov and   Yu.A.Simonov, Phys. Lett. B {\bf 477}, 163 (2000); hep-ph/9912434;  Phys. At. Nucl.  {\bf
63},1428 (2000); hep-ph/9911291; Phys. Lett. B {\bf 636}, 101 (2006); hep-ph/0512151;\\
Yu.A.Simonov, Phys. At. Nucl. {\bf 70}, 44 (2007), hep-ph/0603148.



\bibitem{20}Yu.A.Simonov, Nucl. Phys.  B {\bf 592 }, 350 (2001);
hep-ph/0003114.




\bibitem{21}Yu.A.Simonov, JEPT Lett. {\bf 57}, 513 (1993); Phys.  At. Nucl. {\bf 58 },107
(1995); hep-ph/9311247.


\bibitem{22} A.Yu.Dubin, A.B.Kaidalov and   Yu.A.Simonov, Phys. Lett.  B   {\bf
323}, 41  (1994); Phys. At. Nucl. {\bf 56}, 1745 (1993);   hep-ph/9311344;\\
 Yu.S.Kalashnikova, A.V.Nefediev and Yu.A.Simonov, Phys. Rev. D {\bf  64},
014037 (2001);   hep-ph/0103274.


\bibitem{23} A.M.Badalian, B.L.G.Bakker, Yu.A.Simonov,  Phys.
Rev. {\bf D 75}, 116001 (2007);  hep-ph/0702157.

\bibitem{24} B.Lucini, M.Teper and U.Wenger, JHEP 0406:012, 2004;  hep-lat/0404008.

\bibitem{25}S.M.Fedorov and Yu.A.Simonov, JETP. Lett. {\bf 18}, 57 (2003),
hep-ph/0306216; Yu.A.Simonov, Phys. At. Nucl.  {\bf 67}, 846 (2004);
hep-ph/0302090.




\bibitem{26}Yu.A.Simonov,  Phys.  Lett.  B {\bf 515 }, 137 (2001);
hep-ph/0105141;\\ A.Di Giacomo   and Yu.A.Simonov, Phys. Lett. B  {\bf 595},
368 (2004); hep-ph/0404044.

\bibitem{27} Yu.A.Simonov, JETP Lett. {\bf 54}, 249 (1991); ibid, {\bf 55}, 627 (1992); Phys. Atom. Nucl. {\bf 58} , 309 (1995);  hep-ph/9311216.
\\ H.~G.~Dosch, H.J.Pirner and  Yu.~A.~Simonov, Phys. Lett. B. {\bf 349}, 335
(1995).
\bibitem{28}
Yu.A.Simonov and  M.A.Trusov, Phys. Lett. B {\bf 650}, 36 (2007); arXiv:
0703277 [hep-ph].
\bibitem{29} A.V.Nefediev, Yu.A.Simonov, M.A.Trusov, Int. J. Mod. Phys. {\bf E
18}, 549 (2009).

\bibitem{30}  M.D'Elia, A.Di Giacomo, and   E.Meggiolaro, Phys. Lett B {\bf 408}, 315 (1997); Phys. Rev. D {\bf
  67},  114504 (2003);\\ A.Di Giacomo, E.Meggiolaro  and H.Panagopoulos,
 Nucl. Phys.  B {\bf 483}, 371 (1997).

 \bibitem{31} V.I.Shevchenko and Yu.A.Simonov, Adv. HEPh, 87305 (2009);
 arXiv: 0902.1405 [hep-ph].

 % Phys. Rev. Lett. {\bf 85}, 1811
%(2000), Int. J. Mod, Phys. {\bf A 18}, 127 (2003).

 Yu.A.Simonov,  Trudy Mat. Inst. im. V.A.Steklova  {\bf 272}, 2344 (2010), arXiv: 1003.3608 [hep-ph].


\bibitem{32} Yu.A.Simonov, Phys. Lett. B {\bf 619}, 293 (2005); hep-ph/0502078.

\bibitem{33}  A.Di Giacomo,    E.Meggiolaro, Yu.A.Simonov and A.I.Veselov, Phys. At.
Nucl. {\bf 70},  908  (2007), hep-ph/0512125.

\bibitem{34} M.A.Shifman, A.I.Vainshtein, and V.I.Zakharov, Nucl. Phys.  {\bf B 147}, 385
(1979);\\  B.L.Ioffe,  Prog. Part. Nucl. Phys. {\bf 56}, 232 (2006);
hep-ph/0502148.

\bibitem{35} O.Kaczmarek, F.Karsch, P.Petreczky and F.Zantov, Phys. Lett. B {\bf
543}, 41 (2002); hep-lat/0207002.


\bibitem{36}F.Karsch, Nucl. Phys. B (Proc. Suppl) {\bf 83}, 14 (2000); F.Karsch, E.Laermann and A.Peikert,  hep-lat/0012023.

\bibitem{37}
A.Bazavov, T.Bhattacharya, M. Cheng, C.DeTar, et al., arXiv: 1111.1710
[hep-lat].


\bibitem{38} S.Borsanyi, Z.Fodor, C.Hoelbling, S.D.Katz et al., JHEP 1009.073
(2009); arXiv: 1005. 3508 [hep-lat].
\bibitem{39}P.Petreczky, arXiv: 1001.5284 [hep-ph].
\bibitem{40} A.V.Nefediev and Yu.A.Simonov, Phys.  Rev.  D {\bf 76} , 074014,
(2007); arXiv: 0708.3603 [hep-ph].


\end{thebibliography}
\end{document}